\documentclass[a4paper,fleqn]{cas-dc}

\usepackage[authoryear,longnamesfirst]{natbib}

\def\tsc#1{\csdef{#1}{\textsc{\lowercase{#1}}\xspace}}
\tsc{WGM}
\tsc{QE}
\begin{document}
\let\WriteBookmarks\relax
\def\floatpagepagefraction{1}
\def\textpagefraction{.001}

\shorttitle{CRIS: Collaborative Refinement Integrated with Segmentation Refinement}    

\shortauthors{A. Arudkar, B. Evans}  

\title [mode = title]{CRIS: Collaborative Refinement Integrated with Segmentation for Polyp Segmentation}  

\tnotemark[1] 

\tnotetext[1]{University of Adelaide} 


\author[1]{Ankush Gajanan Arudkar}[
    orcid=0009-0002-3417-7878
]

\cormark[1]

% \fnmark[<footnote mark no>]

\ead{ankushgajanan.arudkar@adelaide.edu.au}

% \ead[url]{<URL>}

% \credit{<Credit authorship details>}

\affiliation[1]{organization={University of Adelaide},
            addressline={Ingkarni Wardli, North Terrace}, 
            city={Adelaide},
%          citysep={}, % Uncomment if no comma needed between city and postcode
            postcode={5000}, 
            state={South Australia},
            country={Australia}}

\author[1]{Bernard J.E. Evans}[
orcid=0000-0002-3517-3775
]

% Footnote of the second author

% Email id of the second author
\ead{bernard.evans@adelaide.edu.au}



% \affiliation[<aff no>]{organization={},
% %          citysep={}, % Uncomment if no comma needed between city and postcode

\cortext[1]{Corresponding author}

\fntext[1]{}


\begin{abstract}
Accurate detection of colorectal cancer and early prevention heavily rely on precise polyp identification during gastrointestinal colonoscopy. Due to limited data, many current state-of-the-art deep learning methods for polyp segmentation often rely on post-processing of masks to reduce noise and enhance results. In this study, we propose an approach that integrates mask refinement and binary semantic segmentation, leveraging a novel collaborative training strategy that surpasses current widely-used refinement strategies. We demonstrate the superiority of our approach through comprehensive evaluation on established benchmark datasets and its successful application across various medical image segmentation architectures.
\end{abstract}


% \begin{highlights}
% \item Introduction of a novel adaptive strategy for refining segmentation masks using end-to-end deep
% learning models.
% \item Proposal of a novel alternating multi-loss training strategy to harness optimization from
% multiple loss functions without experiencing gradient collapse.
% \item Extensive empirical evaluation on two benchmark polyp datasets, comparing the
% segmentation performance of four variants of three medical image segmentation models
% including the proposed strategy.
% \end{highlights}

% Keywords

\maketitle

% Main text
\section{Introduction}
Colorectal cancer (CRC) ranks third globally in cancer prevalence, imposing a significant public health burden \citep{morgan2023global}. Early diagnosis and treatment are pivotal for improving patient outcomes \citep{mahasneh2017molecular}, with timely screening playing a crucial role. Colonoscopy, the widely accepted gold standard for CRC screening, heavily relies on accurate detection of colorectal polyps, which serve as precursors to the disease. While recent advances in image segmentation, driven by machine learning, have significantly enhanced polyp detection capabilities, current deep learning-based methods encounter challenges due to limited labeled datasets. This scarcity often leads to convergence issues, resulting in inaccurate and noisy segmentation masks \citep{liu2021review}. To overcome these limitations, existing methods often incorporate task-specific processing or post-processing steps for mask refinement \citep{wang2011modified,zheng2015conditional,larrazabal2020post}. However, these approaches require careful balancing of precision and recall, with previous studies indicating a tendency to favor one attribute over the other, thereby hindering overall performance \citep{li2024simultaneous}.

We introduce CRIS (Collaborative Refinement Integrated with Segmentation), a novel training strategy designed to enhance polyp segmentation. By simultaneously training a segmentation network and a dedicated mask refinement step, CRIS outperforms conventional deep learning models. This integrated approach, agnostic to backbone architecture, employs interleaved multiple loss training criteria for balanced training. Our empirical study on standard benchmark datasets demonstrates improved precision and recall in polyp detection with this collaborative learning scheme.

The key contributions of this paper are:
\begin{enumerate}
    \item Introducing a novel, integrated approach for training binary semantic image segmentation models with a trainable mask refinement step.
    \item Proposing an interleaved dual-loss learning strategy to effectively balance gradients from segmentation and refinement objectives.
    \item Conducting an empirical study on two benchmark datasets, CVC-ClinicDB and Kvasir-SEG, utilizing various image segmentation architectures with both existing and newly proposed mask refinement techniques.
\end{enumerate}

\section{Related work}

In recent years, deep learning methods, including specialized networks like PolypSegNet and PraNet, have dominated the domain of semantic polyp segmentation \citep{akbari2018polyp,brandao2017fully,mahmud2021polypsegnet,fan2020pranet}. Traditionally, post-processing steps were commonly employed to refine segmentation masks, treating refinement as a separate process from the network itself. This paper introduces a unified strategy that integrates end-to-end learning and a refining network for the binary semantic segmentation of polyps within a single training process.

\textbf{Polyp segmentation:} Traditional models for medical image segmentation, such as U-Net and U-Net++ \citep{unetronneberger2015u,unet++zhou2018unet++}, have found extensive application, with variants designed specifically for polyp segmentation. Recent approaches leverage transfer learning \citep{wen2023rethinking} to fine-tune existing models, including large foundational models like the Segment Anything Model \citep{huang2024segment}. However, these approaches are constrained by data availability and the inherent bias of foundational models toward natural images, with larger models yielding inequitable results compared to their size.

\textbf{Post-processing mask refinement:} In data-constrained segmentation tasks, mask refinement has often relied on image processing techniques, one example being watershed transforms modified for medical images \citep{wang2011modified}. Recent methods have explored the use of additional networks to refine the output produced by the base model \citep{larrazabal2020post,kim2021model}. These approaches enhance the overall quality of segmentation masks independently of the base model's training.

\textbf{Adaptive mask refinement:} Conditional random fields have been employed to refine segmentation outputs \citep{zheng2015conditional,vemulapalli2016gaussian}, serving as the final stage of many segmentation models with trainable parameters. End-to-end networks have also been utilized for this task, leveraging features generated by a backbone encoder with a trainable decoder network \citep{nguyen2020contour}. This approach shares decoding and refinement steps. Noteworthy strategies, including those tested alongside the proposed methods, are discussed in the main evaluation.

\section{Methodology}
% introducing paragraph
% 1. Problem definition
% 2. Refinement module
% 3. Training
% 4. Metrics

\begin{figure}[!t]
  \centering
  \includegraphics[scale=0.45]{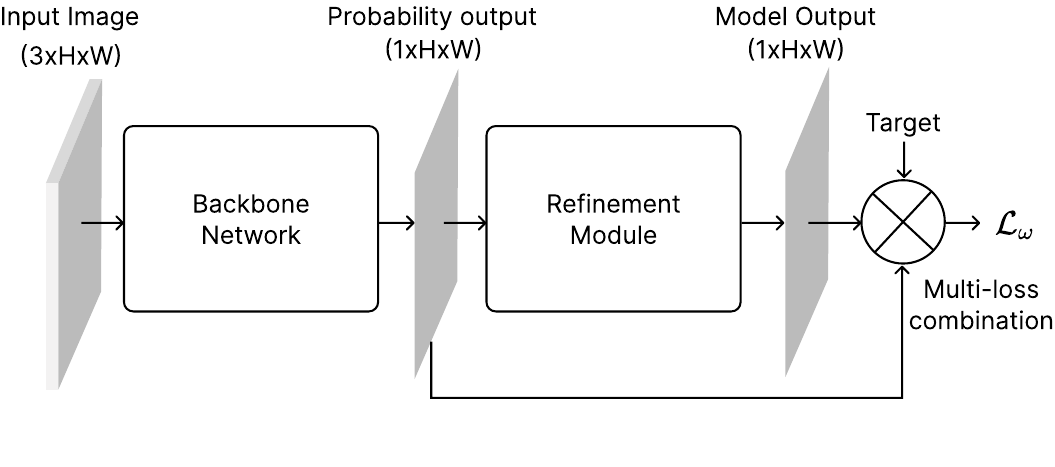}
  \caption{Proposed architecture of the CRIS refinement strategy with backbone network and refinement module producing individual losses trained by proposed interleaved training strategy.}
  \label{fig:arch}
\end{figure}

This paper introduces a refinement strategy designed to enhance semantic binary segmentation masks, seamlessly integrable into any mask-generating backbone network. We provide a comprehensive explanation of the refinement module, followed by a description of the proposed interleaved multi-loss training strategy. We demonstrate its superiority over the simplistic approach of adding additional layers to model. The proposed architecture of the mask refinement strategy is illustrated in Figure \ref{fig:arch}.

\subsection{Datasets}
To assess the efficacy of our approach, we conducted experiments on two widely recognized colonoscopy datasets frequently employed for benchmarking polyp segmentation outcomes:

\begin{enumerate}
    \item \textbf{Kvasir-SEG} \citep{kvasirjha2020kvasir}: Comprising 1000 polyp images, this dataset incorporates ground truth masks obtained from Vestre Viken Health Trust in Norway.

    \item \textbf{CVC-ClinicDB} \citep{cvcvazquez2017benchmark}: Encompassing 612 annotated images derived from 29 video sequences involving 23 patients.
\end{enumerate}

\subsection{Problem definition}
The model is trained on image pairs, $(I^i, G^i) \in \mathcal{D}$ from the dataset $\mathcal{D}$, where the dimensions of $I$ are $3 \times H \times W$, and the ground truth $G$ is of dimensions $1 \times H \times W$, with $H$ representing the height and $W$ the width of the image. Our proposed framework aligns with the settings of binary semantic segmentation. In this context, the input image $I$ undergoes transformation by the model $\mathcal{M}$ to yield an output mask $\mathcal{M}(I)$ of dimensions $1 \times H \times W$ as shown in Figure \ref{fig:arch}. Each element in this output mask denotes the probability of the corresponding pixel belonging to a polyp. We formulate this problem as the minimization of the cumulative loss function $\mathcal{L}_{\omega}$, computed between the ground truth $G^i$ and the model output $\mathcal{M}(I^i)$.

\subsection{Refinement module}

\begin{figure*}[!t]
  \centering
  \includegraphics[width=0.8\textwidth]{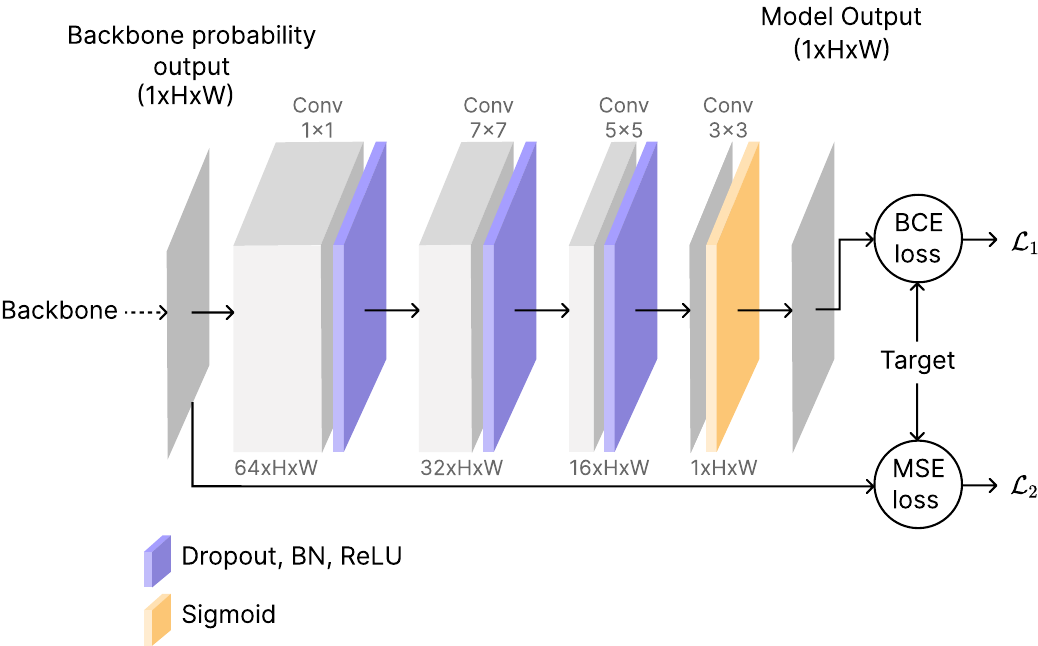}
  \caption{Refinement module network appended to base model for collaborative training with loss from base model and refined output.}
  \label{fig:refine_mod}
\end{figure*}

We propose employing a fully convolutional network (FCN) as the refinement network along with a predetermined backbone network, accompanied by a novel training strategy. This module takes the probability map generated by the backbone network as input and utilizes $1\times1$ convolutions to expand the channel information in the input. As depicted in Figure \ref{fig:refine_mod}, it is then followed by a sequence of convolutional filters with decreasing kernel sizes.

To mitigate the risk of overfitting resulting from the inclusion of additional layers, we incorporate dropout with a probability of 0.01 after each set of convolutional operations between every trainable layer of the module. The output of the refinement module is a probability map indicating the likelihood of polyp presence at each pixel location. The loss is calculated using the intermediate and final masks produced by the model, as elaborated in the subsequent section.

\subsection{Training procedure}
% data splitting strategy
% hypterparameter tuning
% loss function

\noindent\textbf{Data preparation:} The datasets were systematically partitioned into training ($\mathcal{D}_{\text{train}}$), validation ($\mathcal{D}_{\text{val}}$), and test sets ($\mathcal{D}_{\text{test}}$), with proportions of 70\%, 15\%, and 15\%, respectively. This partitioning was conducted using a pseudo-random approach with a fixed seed for each dataset, ensuring consistent training, validation, and testing datasets across different models. Subsequently, each model underwent training on the designated training dataset, and its performance was subsequently evaluated on the respective testing dataset.

\noindent\textbf{Loss function:} To train the entire model we propose a strategy to train the model and refinement module with different loss criteria. We use a combination of mean squared error loss for intermediate backbone output and binary cross entropy for final output of the model. The errors for an image $I^i$ with ground truth $G^i$ and backbone model $\mathcal{B}$ combined with refinement module to for final model $\mathcal{M_B}$ are calculated as follows

\begin{equation}
    \mathcal{L}_{1,i} = \frac{1}{H\times W}\sum_{x=1,y=1}^{H,W} (\mathcal{B}(I^i)_{x,y} - G^i_{x,y})^2
\end{equation}

\begin{equation}
\begin{aligned}
    \mathcal{L}_{2,i} &= \frac{1}{H\times W}\sum_{x=1,y=1}^{H,W}[-G^i_{x,y}\log(\mathcal{M_B}(I^i)_{x,y}) \\
    &\quad+ (1-G^i_{x,y})\log(1-\mathcal{M_B}(I^i)_{x,y})]
\end{aligned}
\end{equation}

By employing a dual loss strategy, the model effectively safeguards against parameter collapse across the backbone and refinement module, ensuring that segmentation masks are learned as distinct entities for both networks. Through $\mathcal{L}_1$, the base model is trained to acquire proficiency in generating semantic masks without tight coupling with refinement layers, while $\mathcal{L}_2$ ensures that the refinement block builds upon the knowledge of the base network to further enhance the final output. In each epoch, the losses were alternately backpropagated rather than computing a weighted sum, preventing the gradients of one loss from overpowering the other, thereby ensuring satisfactory learning of both criteria. The combined loss for epochs with index $e$ is expressed as follows

\begin{equation}
    \mathcal{L}_{\omega} = \sum_{i\epsilon\mathcal{D}}(1-(e\ mod\ 2))\mathcal{L}_{1,i} + (e\ mod\ 2)\mathcal{L}_{2,i}
\end{equation}

\subsection{Metrics}
The Metrics $M(I,\hat{I})$ measure the quality of generated masks $\hat{I}$ and probability maps $\mathcal{M}(I)$ to the ground truth $G$.

\textbf{Precision and Recall:} 
We employ precision and recall curves to assess the quality of predictions made by the model across various thresholds. For a given prediction $\hat{I}$ and ground truth $G$, the metrics are computed as follows:

\begin{equation}
    Precision = \frac{|\hat{I}\cap G|}{|\hat{I}\cap G|+|\hat{I}\setminus G|}
\end{equation}

\begin{equation}
    Recall = \frac{|\hat{I}\cap G|}{|\hat{I}\cap G|+|G\setminus \hat{I}|}
\end{equation}

\textbf{DICE score:} 
To assess the quality of thresholded binary masks, we utilize the DICE score as a comparative metric across different models:

\begin{equation}
DICE = \frac{2.|\hat{I}\cap G|}{|\hat{I}| + |G|}
\end{equation}

\section{Experiments}
% baseline
% baseline+FCN
% crf
% proposed

% introduce
% experimental setting
% Comparision baselines
% - base,  base+FCN, CRF, Proposed
% Comparision results

We evaluate the effectiveness of the proposed refinement strategy on benchmark colonoscopy datasets—CVC-ClinicDB \citep{cvcvazquez2017benchmark} and Kvasir-SEG \citep{kvasirjha2020kvasir}.

\subsection{Comparison baselines}

We evaluated the proposed strategy on established medical image segmentation models—UNet \citep{unetronneberger2015u}, UNet++ \citep{unet++zhou2018unet++}, and SegNet \citep{badrinarayanan2017segnet}. Training from scratch involved an ablation study with two datasets and four strategies:

\begin{enumerate}
    \item \textbf{Baseline:} Unmodified version of the base architecture.
    \item \textbf{Conditional Random Field (CRF):} Base models were trained with the state-of-the-art CRF refinement strategy \citep{zheng2015conditional} for segmentation masks.
    \item \textbf{Backbone+FCN:} The base model augmented with FCN layers of the refinement block, trained without the proposed training method.
    \item \textbf{Proposed Strategy:} Augmentation of the base model with a refinement block, utilizing proposed collaborative training strategy.
\end{enumerate}

\subsection{Experimental settings}
We conducted an extensive study involving a combination of two benchmark datasets, three choices for the base model, and four refinement strategies for each base model. Each model received the same versions of training, validation, and testing datasets, and model hyperparameters were tuned using the strategy outlined in \citep{akiba2019optuna}, with 20 trials for each combination. The assessment involved a comparison of the quality of masks generated by probability masks as well as thresholded masks. The thresholds for each model were determined through grid search to maximize the DICE score on the training dataset.

For the experiments, we trained all models on an AMD Ryzen 9 3900X 12-Core Processor CPU with an NVIDIA GeForce RTX2070 GPU. A consistent batch size of 4 was used for training, and the tuning of learning rates was subject to pruning via the median pruning strategy within the range of 1e-2 to 1e-6. The Adam optimizer was employed for training the models with betas as (0.9, 0.999), with other hyperparameters held constant.

\subsection{Comparison results}

\textbf{Quantitative metrics:}
We present a quantitative comparison in Table \ref{tab:kv} for the Kvasir-SEG dataset, where the proposed strategy showcases superior performance. Metrics were computed using the best threshold identified on the training dataset. Notably, the proposed refinement module outperforms models with similar network structures. Our proposed training method is a plausible explanation for this superiority, preventing the training of the base network and refinement module from collapsing. This approach enables the base network to acquire meaningful mask representations before the refinement module refines the produced masks.

% kvasir
% \setlength{\tabcolsep}{4pt}
\begin{table}[!t]
\caption{DICE score and Mean squared error for Backbone models (UNet, UNet++, SegNet) with baseline, state-of-the-art and proposed refinement strategy on Kvasir-SEG dataset.}
\label{tab:kv}
% \vskip 0.005in
\begin{center}
\begin{small}
% \begin{sc}
\resizebox{0.48\textwidth}{!}{%
\begin{tabular}{|c|rr|rr|rr|}
\hline
\multirow{2}{*}{Strategy} & \multicolumn{2}{c|}{UNet}                             & \multicolumn{2}{c|}{UNet++}                           & \multicolumn{2}{c|}{SegNet}                           \\ \cline{2-7} 
                          & \multicolumn{1}{c|}{DICE}  & \multicolumn{1}{c|}{MSE} & \multicolumn{1}{c|}{DICE}  & \multicolumn{1}{c|}{MSE} & \multicolumn{1}{c|}{DICE}  & \multicolumn{1}{c|}{MSE} \\ \hline
Backbone                      & \multicolumn{1}{r|}{76.69} & 0.049                    & \multicolumn{1}{r|}{83.08} & 0.043                    & \multicolumn{1}{r|}{71.54} & 0.060                    \\
Backbone+FCN                  & \multicolumn{1}{r|}{85.46} & \textbf{0.034}                    & \multicolumn{1}{r|}{84.35} & 0.034                    & \multicolumn{1}{r|}{73.88} & 0.052                    \\
Backbone+CRF                  & \multicolumn{1}{r|}{84.41} & 0.038                    & \multicolumn{1}{r|}{82.57} & 0.037                    & \multicolumn{1}{r|}{83.01} & 0.042                    \\
\textbf{Proposed}                  & \multicolumn{1}{r|}{\textbf{85.99}} & 0.035                    & \multicolumn{1}{r|}{\textbf{86.05}} & \textbf{0.032}                    & \multicolumn{1}{r|}{\textbf{86.88}} & \textbf{0.041}                    \\ \hline
\end{tabular}
% \end{sc}
}
\end{small}
\end{center}
\vskip -0.1in
\end{table}

\begin{figure*}[!hbt]
  \centering
  \includegraphics[width=\textwidth]{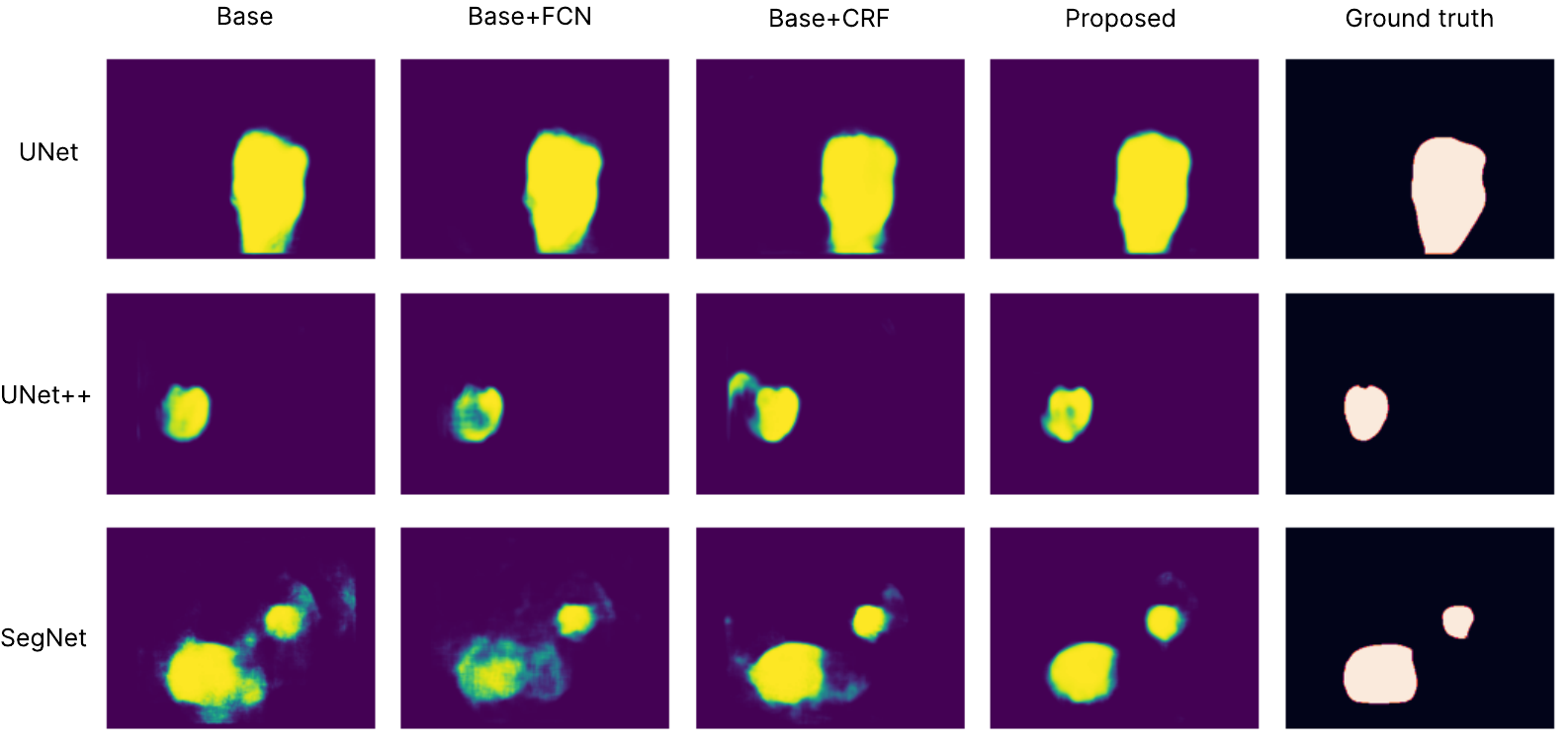}
  \caption{Probability attention maps generated by combination of backbone architectures (rows) with refinement strategies used (columns) for test images of CVC-ClinicDB \citep{cvcvazquez2017benchmark} dataset.}
  \label{fig:results}
\end{figure*}

Similar trends are observed for the CVC-ClinicDB dataset, where the proposed method consistently outperforms most other models. Our method surpasses both the base models and models with similar network structures. It is essential to highlight that, for a fair comparison, the same FCN structure for refinement modules was maintained across all networks, even in scenarios where different layer structures for the refinement block could be advantageous. Reduced noise in probability maps generated by the proposed method is evident across various models, as illustrated in Figure \ref{fig:results}.

% cvc
\begin{table}[]
\caption{DICE score and Mean squared error for Backbone models (UNet, UNet++, SegNet) with baseline, state-of-the-art and proposed refinement strategy on CVC-ClinicDB dataset.}\label{tab:cvc}
\begin{center}
\begin{small}
% \begin{sc}
\resizebox{0.48\textwidth}{!}{%
\begin{tabular}{|c|cr|cr|cr|}
\hline
\multirow{2}{*}{Strategy} & \multicolumn{2}{c|}{UNet}                               & \multicolumn{2}{c|}{UNet++}                             & \multicolumn{2}{c|}{SegNet}                             \\ \cline{2-7} 
                          & \multicolumn{1}{c|}{DICE}  & \multicolumn{1}{c|}{MSE}   & \multicolumn{1}{c|}{DICE}  & \multicolumn{1}{c|}{MSE}   & \multicolumn{1}{c|}{DICE}  & \multicolumn{1}{c|}{MSE}   \\ \hline
Backbone                      & \multicolumn{1}{r|}{91.58} & 0.011                      & \multicolumn{1}{r|}{91.25} & 0.011                      & \multicolumn{1}{r|}{86.39} & 0.020                      \\
Backbone+FCN                  & \multicolumn{1}{c|}{91.68} & \multicolumn{1}{c|}{0.010} & \multicolumn{1}{c|}{90.94} & \multicolumn{1}{c|}{0.011} & \multicolumn{1}{c|}{78.54} & \multicolumn{1}{c|}{0.022} \\
Backbone+CRF                  & \multicolumn{1}{r|}{88.82} & 0.013                      & \multicolumn{1}{r|}{\textbf{92.81}} & 0.010                      & \multicolumn{1}{r|}{90.15} & 0.011                      \\
\textbf{Proposed}                  & \multicolumn{1}{r|}{\textbf{92.62}} & \textbf{0.010}                      & \multicolumn{1}{r|}{92.49} & \textbf{0.009}                      & \multicolumn{1}{r|}{\textbf{92.39}} & \textbf{0.008}                      \\ \hline
\end{tabular}
% \end{sc}
}
\end{small}
\end{center}
\end{table}

\begin{figure*}[hbt!]
  \centering
  \includegraphics[width=\textwidth]{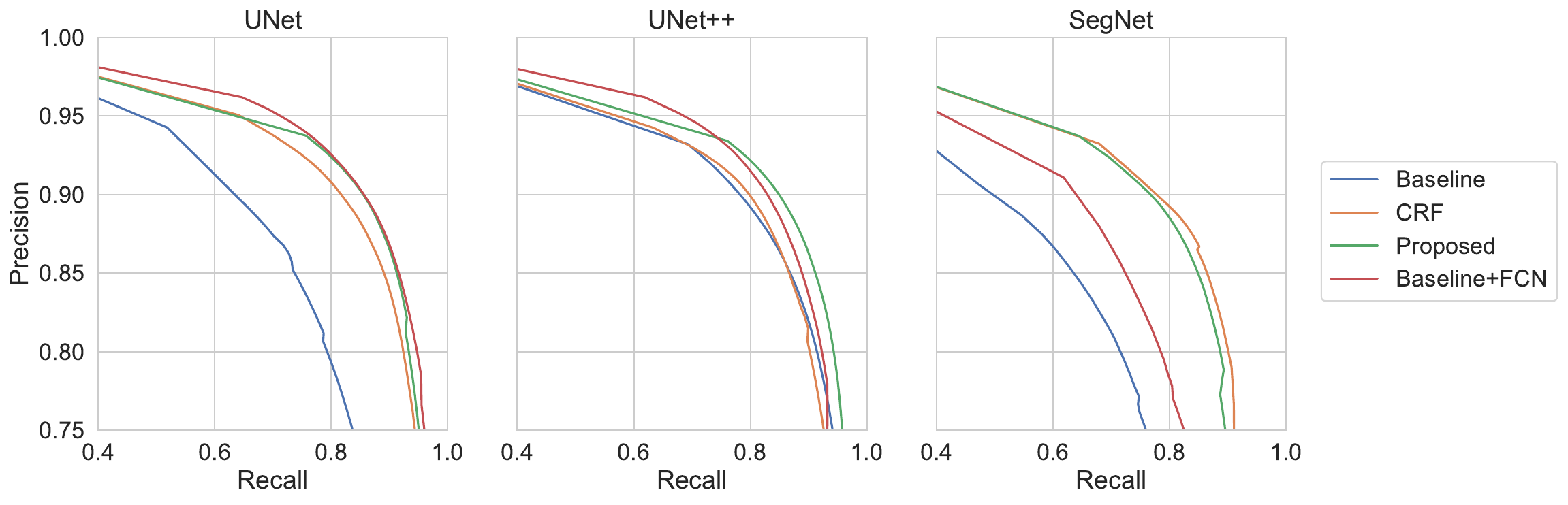}
  \caption{Precision-Recall curves for considered model architectures and refinement strategy for Kvasir-SEG \citep{kvasirjha2020kvasir} dataset, highlighting the consistent better performance of proposed method over other models.}
  \label{fig:kv-pr}
\end{figure*}

\noindent\textbf{Precision-Recall curves:} In Figure \ref{fig:kv-pr}, we present the precision-recall curve for models trained on the Kvasir-SEG dataset. Notably, the proposed refinement method consistently outperforms both the base model and the base model with FCN layers. We observed that, for most points along the precision-recall trade-off curve, the proposed method yields superior results compared to models incorporating Conditional Random Fields (CRF).

\begin{figure*}[hbt!]
  \centering
  \includegraphics[width=\textwidth]{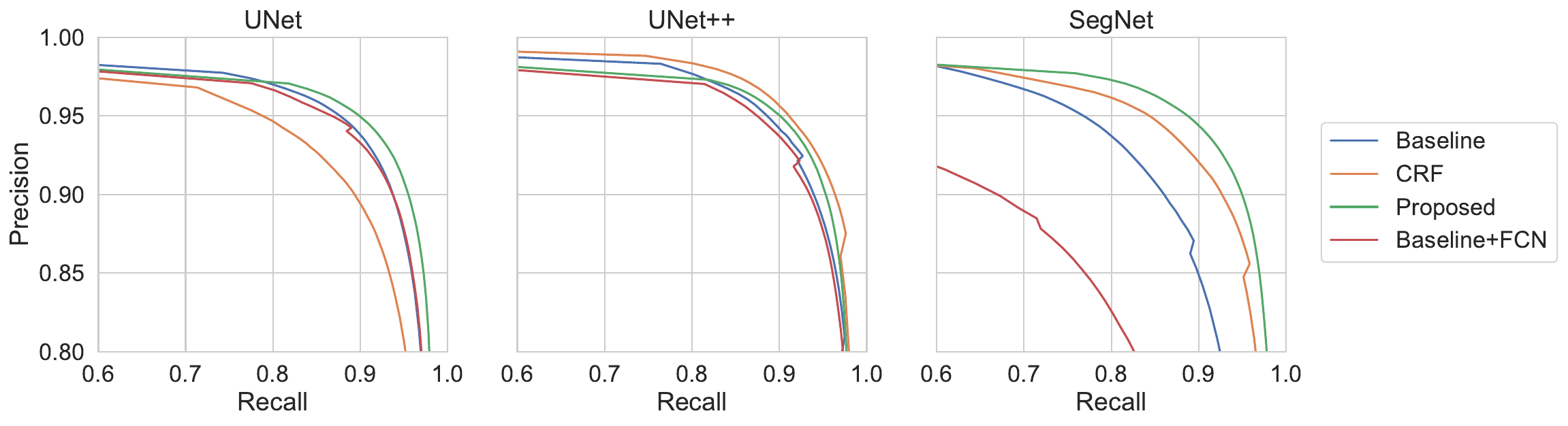}
  \caption{Precision-Recall curves for considered model architectures and refinement strategy for CVC-ClinicDB \citep{cvcvazquez2017benchmark} dataset.}
  \label{fig:cvc-pr}
\end{figure*}

In Figure \ref{fig:cvc-pr}, the precision-recall curves for models trained on the CVC-ClinicDB dataset are depicted. Once again, the proposed method showcases superior performance compared to both the base models and the base models with FCN layers. An interesting observation is the varied performance of Conditional Random Fields (CRF), which appears to depend on the underlying architecture. In particular, it performs worse than the base model for UNet and outperforms all models for UNet++. Conversely, our proposed method demonstrates more consistent behavior across the precision-recall trade-off.

\section{Conclusion}
In conclusion, our proposed Collaborative Refinement Integrated with Segmentation (CRIS) strategy significantly enhanced semantic binary segmentation masks for colorectal polyp detection during colonoscopy. Through extensive experimentation on Kvasir-SEG and CVC-ClinicDB datasets, our approach consistently outperformed state-of-the-art models and refinement strategies. The interleaved loss propagation across layers ensures a complementary evolution of the base network and refinement module, leading to improved overall segmentation accuracy. Our findings bear significance for advancing colorectal polyp detection, promising improved clinical outcomes through accurate and reliable medical image segmentation. In future, we intend to explore further refinement and broader applications of our strategy in multi-class medical image segmentation with varying network architectures in refinement modules.

% Use the style of numbering in square brackets.
% If nothing is used, default style will be taken.
%\begin{enumerate}[a)]
%\item 
%\item 
%\item 
%\end{enumerate}  

%\begin{itemize}
%\item 
%\item 
%\item 
%\end{itemize}  

%\begin{description}
%\item[]
%\item[] 
%\item[] 
%\end{description}  

% Figure
% \begin{figure}[<options>]
% 	\centering
% 		\includegraphics[<options>]{}
% 	  \caption{}\label{fig1}
% \end{figure}

% \begin{table}[<options>]
% \begin{tabular*}{\tblwidth}{@{}LL@{}}
% \toprule
% \midrule
%  & \\
%  & \\
%  & \\
%  & \\
% \bottomrule
% \end{tabular*}
% \end{table}

% Uncomment and use as the case may be
%\begin{theorem} 
%\end{theorem}

% Uncomment and use as the case may be
%\begin{lemma} 
%\end{lemma}


% \section{}\label{}

\printcredits

\bibliographystyle{cas-model2-names}


\bibliography{cas-refs}

% % Here goes the biography details.

% % Here goes the biography details.

\end{document}